\documentclass{sigchi}

\usepackage{times}
\usepackage{helvet}
\usepackage{courier}
\usepackage{balance}

\usepackage[utf8]{inputenc}
\usepackage{url}
\makeatletter
\def\url@leostyle{%
  \@ifundefined{selectfont}{\def\UrlFont{\sf}}{\def\UrlFont{\small\bf\ttfamily}}}
\makeatother

\def\pprw{8.5in}
\def\pprh{11in}

\usepackage{microtype}
\usepackage{graphicx}
\usepackage{booktabs}
\usepackage{tabulary}
\usepackage{subcaption}
\usepackage{flushend}
\usepackage{amsmath}

\usepackage{enumitem}
\setlist{nolistsep}

\usepackage{algorithmicx}
\usepackage{algorithm}
\usepackage{algpseudocode}

\newfont{\mycrnotice}{ptmr8t at 7pt}
\newfont{\myconfname}{ptmri8t at 7pt}
\let\crnotice\mycrnotice%
\let\confname\myconfname%

\newcommand{\ie}{\emph{i.\,e.}}

\newcommand{\eg}{\emph{e.\,g.}}
 
\newcommand{\wrt}{with respect to }

\usepackage{gensymb}

\newcommand{\spara}[1]{\smallskip\noindent{\bf #1.}}

\usepackage{ifthen}
\newboolean{usebiber}
\setboolean{usebiber}{false}

\ifthenelse{\boolean{usebiber}}{\usepackage[backend=biber,natbib=true,style=sigchi,maxcitenames=1,maxbibnames=100]{biblatex}}{\usepackage[backend=bibtex,natbib=true,style=sigchi,maxcitenames=1,maxbibnames=100]{biblatex}}

\renewcommand{\cite}{\citep}

\usepackage{xpatch}

\xpatchbibmacro{name:andothers}{%
  \bibstring{andothers}%
}{%
  \bibstring[\emph]{andothers}%
}{}{}

\usepackage[draft]{hyperref}

\numberofauthors{3}

\author{
\alignauthor
Eduardo Graells-Garrido\thanks{Corresponding author: \url{eduardo.graells@telefonica.com}. Work primarily carried out while the first author was a PhD student in the Web Research Group, at Universitat Pompeu Fabra, Barcelona, Spain.} \\
\affaddr{Telefónica I+D} \\
\affaddr{Santiago, Chile} 
\alignauthor
Mounia Lalmas \\
\affaddr{Yahoo Labs} \\
\affaddr{London, UK} 
\alignauthor
Ricardo Baeza-Yates \\
\affaddr{Yahoo Labs} \\
\affaddr{Sunnyvale, USA} 
}

\title{Data Portraits and Intermediary Topics: \\ Encouraging Exploration of Politically Diverse Profiles}

\toappear{This is an author generated version. Paper to be presented at ACM Intelligent User Interfaces (IUI 2016). The final version can be obtained at DOI: http://dx.doi.org/10.1145/2856767.2856776}

\begin{document}

\maketitle

\begin{abstract} 
In micro-blogging platforms, people connect and interact with others.
However, due to cognitive biases, they tend to interact with like-minded people and read agreeable information only.
Many efforts to make people connect with those who think differently have not worked well. 
In this paper, we hypothesize, first, that previous approaches have not worked because they have been direct -- they have tried to explicitly connect people with those having opposing views on sensitive issues. Second, that neither recommendation or presentation of information by themselves are enough to encourage behavioral change.
We propose a platform that mixes a recommender algorithm and a visualization-based user interface to explore recommendations.
It recommends politically diverse profiles in terms of distance of latent topics, and displays those recommendations in a visual representation of each user's personal content.
We performed an ``in the wild'' evaluation of this platform, and found that people explored more recommendations when using a biased algorithm instead of ours. In line with our hypothesis, we also found that the mixture of our recommender algorithm and our user interface, allowed politically interested users to exhibit an unbiased exploration of the recommended profiles. 
Finally, our results contribute insights in two aspects: first, which individual differences are important when designing platforms aimed at behavioral change; and second, which  algorithms and user interfaces should be mixed to help users avoid cognitive mechanisms that lead to biased behavior.
\end{abstract}
 
\sloppy 

 
\section{Introduction}
Research from social sciences has shown that, while everyone seems to have a voice on the Web, people tend to listen and connect to those with similar beliefs in political and ideological issues. 
Such behavior can be explained in terms of homophily~\cite{mcpherson2001birds}, a cognitive bias.
Homophily is present in many situations and can be beneficial, as communication with culturally alike people is easier to handle. 
However, \wrt ideological issues it can have serious consequences, both off- and on-line.
On one hand, groups of like-minded users tend to disconnect from other groups, polarizing group views~\cite{JOPP:JOPP148}. 
On the other hand, Web platforms recommend and adapt content based on interaction and network data of users, \ie, who is connected to them and what they have liked before. 
Because one of their main goals is to maximize user engagement, recommendation algorithms often push content that reinforces homophily in behavior by displaying mostly agreeable information.
Such biased reinforcement, in turn, makes further recommendations consisting of even more polarized content, confining users to \emph{filter bubbles} \cite{pariser2011filter}. 

Until now, most approaches have focused on how to motivate users to read challenging information or how to motivate a change in behavior through recommender systems and display of potentially challenging information. 
This direct approach has not been effective as users often do not value diversity or do not feel satisfied with it \cite{munson2010presenting}.
Motivated by this scenario, our work aims at understanding how to encourage exposure to diverse people from an ideological point of view on micro-blogging platforms. Our research questions are:

\begin{center}
Is it possible to encourage exploration and acceptance of user profiles recommended on the basis of political diversity? 
If so, which factors influence this behavior?
\end{center}

We propose an indirect approach using an advanced user interface, because we believe that neither a recommender system nor a user interface alone are enough to encourage unbiased behavior. Our hypothesis is that an indirect, mixed approach helps users to overcome the cognitive dissonance produced by exposure to potentially challenging information.

To achieve this mixed approach, our work combines the output of a recommender algorithm with a visual depiction. The recommender algorithm is aimed at recommending people who may think differently based on a proxy latent space modeling of topics, named \textit{intermediary topics}~\cite{graells2014people}. Intermediary topics are latent topics that users, with distant views on some sensitive issues, have in common (\eg, a music genre, a cuisine type). These topics are our means to introduce users to each other, knowing that first impressions matter \cite{asch1946forming}, thus giving connection a chance to happen.

The visual depiction is aimed, first, at providing a pleasing and joyful view of the user's own interests, and then to provide a visual display of recommended profiles.
Particularly, our proposal is a visualization of user profiles called \emph{data portrait}~\cite{donath2010data}, with the purpose of making users aware of their own interests and the image they project on a social platform.
The visualization of recommendations -- of people to ``follow'' -- is based on a hierarchical visualization technique displaying how \textit{recommendees} can be grouped.
Both concepts allow users to receive recommendations in the context of their data portrait.

Our system was deployed ``in the wild'' to evaluate its effects on users, using a proof-of-concept implementation of a recommender system based on the intermediary topics paradigm.
We focused on Chile, a Latin-American country with one of the highest Internet penetration rates among developing countries~\cite{pewresearch2015}, and whose population actively discuss politics on Twitter~\cite{valenzuela2014facebook}. As found in a previous study, such discussion is homophilic \cite{DBLP:journals/corr/Graells-Garrido15a}, making it a good candidate for analysis.
We analyzed how Chilean users interact with their portraits and explore the recommendations within the realm of \textit{Casual Information Visualization} \cite{pousman2007casual}. We did not frame our evaluation using specific tasks to be performed by users, as these would not reflect scenarios for our system. 
Instead, we analyzed user behavior employing user engagement metrics \cite{lalmas2014measuring}, as these are more appropriate for studying non-goal oriented user behaviors. 
The main results of our experiment are:

\begin{itemize}
\item The usage of visualization to depict recommendations alongside a data portrait allowed users to explore more recommended profiles, regardless of the recommender algorithm used to generate them.
\item Behavioral differences in terms of political involvement influenced how users interacted and engaged with our system.
\item On a standalone basis, neither recommender system or visualization help users to exhibit a conscious (unbiased) exploration. However, their combination is effective when users have political content in their profiles. 
\end{itemize}

These results have implications on the design of systems aimed at exploring user generated content: a one-size-fits-all approach misses the opportunity of giving users tools to get the best out of their exploring experience. 
We discuss design implications in terms of who can be targeted with systems like ours and when to consider presenting diverse recommendations. Our main conclusion is that indirect approaches like these data portraits can help users to make conscious decisions in these biased scenarios.

\section{Related Work}

\subsection{Homophily and Content Recommendation}
\textit{Homophily} is the tendency to form ties with similar others, where similarity can be bound to factors ranging from socio-demographic to behavioral and intra-personal ones~\citep{mcpherson2001birds}.
In micro-blogging platforms, the presence of homophily in how individuals interact has been shown to be reflected in the structure of their ego-network, which has allowed to predict user attributes \cite{al2012homophily,rout2013s}, including political leaning \cite{barbera2015birds}.
Likewise, its presence has been used to recommend people to interact with \cite{chen2009make,hannon2010recommending}. 

In this paper, we focus on recommendations on micro-blogging platforms based on user similarity. Similarity can be defined in different ways.
For instance, two users might be similar if they use the same \textit{hashtags} \cite{brzozowski2011should}, follow the same accounts \cite{goel2013discovering}, mention the same entities \cite{michelson2010discovering}, or have similar \textit{latent topics} \cite{ramage2010characterizing} as estimated using Latent Dirichlet Allocation (LDA hereafter)~\cite{blei2003latent}.
However, similarity is not the only feature to consider when recommending information or people to follow. 
Other features include content quality and popularity \cite{chen2012collaborative}, network relevance (\textit{friend of a friend})~\cite{chen2009make}, explainability \cite{herlocker2000explaining}, and centrality measures~\cite{gupta2013wtf}. 

In our work, we propose \textit{intermediary topics} \cite{graells2014people} as a feature to consider when recommending users to follow. 
The intuition behind intermediary topics is that they focus on homophily on the basis of specific shared latent topics computed using LDA \cite{blei2003latent}. LDA has been found to be reliable for user classification~\cite{pennacchiotti2011democrats}, and in our context, it has successfully identified latent topics that could act as intermediary between people with diverse political profiles~\cite{DBLP:journals/corr/Graells-Garrido15a}.

Using topics and relationships between them, represented through the so-called \textit{topic graphs}, for recommendation purposes is not new. 
For instance, \citet{gretarsson2012topicnets} visualize topic graphs to ease knowledge discovery. Our context is different. We do not attempt to visualize them;  instead, we use the intermediary topics as input features for generating the recommendations. We nonetheless experiment with visual depiction of recommendations, through the paradigm of data portraits, because the visualization of social recommendations has been shown to boost user satisfaction~\cite{gretarsson2010smallworlds}.

\subsection{Encouraging Exposure to Diverse Content}
Exposure to exclusively agreeable information or like-minded people reinforces and polarizes individual and group stances on ideological issues~\cite{myers1975polarizing,sunstein2009going}.
Various works have looked into ways to improve the exposure to challenging information, by employing algorithms for content selection, as well as changing the depiction of this type of information. The degrees of success vary, but results have not been as good as expected in terms of behavioral change. 

\emph{Opinion Space} \cite{faridani2010opinion} is a self-organizing interactive visualization of an information space. In it, individual opinions of participants in debate were visualized according to their \textit{opinion profiles}, built automatically for each participant after answering questions about key political issues. 
Although the visual approach did not reduce selective exposure, it generated more engagement than baseline text-based interfaces and users were more respectful with those having opposite opinions.
In \emph{NewsCube}~\cite{park2009newscube}, several automatically determined aspects of news stories in political contexts are presented to allow users to access diverse points of view of news events in political contexts. Each aspect is displayed in its own cluster, allowing users to see the diversity of available points of view. This clustered presentation augments the number of interactions with news, but not the number of interactions with different, opposing, clusters \cite{chhabra2013does}.
\citet{munson2010presenting} tested different ways of altering a user interface without changing its core interaction mechanisms, by changing the sorting order of information and highlighting items pertaining to opposite points of view \wrt the user. It was found that only a minority of users, called \textit{diversity-aware}, values diversity.
In discussion forums, \citet{liao2014can} added \textit{position indicators} of stance polarization to participants, improving agreement of users with those of opposing views when their positions were not consistently moderate, or when the information seekers were looking for highly accurate information.

One possible reason of why direct approaches have not worked is the \textit{selective exposure}~\cite{hart2009feeling} mechanism. 
Selective exposure makes users discard potentially challenging information, either by not reading it, not accepting the recommendation, or preferring an agreeable alternative, regardless of the factual value of the agreeable information item. Through selective exposure, users avoid \emph{cognitive dissonance} \cite{festinger1962theory}, a state of discomfort that affects persons confronted with conflicting ideas, beliefs, values or emotional reactions. Thus, if users were to receive recommendations of users who think differently, the selective exposure mechanism would likely prevent them for even browsing the recommendations.

In our work, we apply an indirect approach, where we use intermediary topics \cite{graells2014people} to recommend people with potentially distant opinions (as captured by LDA), and yet with specific interests in common. 
In contrast to the approaches above mentioned, the recommendations are not based on their (non)-alignment on politics or sensitive issues. Instead, we build a data portrait of users of micro-blogging platforms, and show recommendations in that context, emphasizing the similarity of users in terms of their alignment on intermediary topics.

\subsection{Information Visualization}
Our work is related to the field known as \emph{Casual Information Visualization}, defined by \citet{pousman2007casual} as \emph{``the use of computer mediated tools to depict personally meaningful information in visual ways that support everyday users in both everyday work and non-work situations.''}
The focus on everyday situations means that there does not need to be a concrete task to be completed, nor a specific analytic insight to be expected.

\citet{yi2008understanding} identify four cognitive processes that lead to insights gained through visualization: \textit{provide overview}, \textit{adjust}, \textit{detect pattern}, and \textit{match mental model}. Those relevant to our context are \textit{provide overview} and \textit{match mental model}. 
To provide an overview of profiles and match mental models of portrayed users, we use \textit{word clouds} as the primary element of our proposed visualization.
Word clouds have a long history in information visualization \cite{viegas2008timelines}. Although not appropriate for analytical tasks, word clouds  are expressive, familiar, and popular with users, as they help them expressing themselves~\cite{viegas2009participatory}. 
We use word clouds both to provide an overview of a profile and as a navigational tool to explore it, in a coordinated view with other visual elements of the portrait.

Other visualization techniques such as \textit{WordTrees}~\cite{wattenberg2008word} and \textit{PhraseNets}~\cite{van2009mapping} have been employed to depict structure in text. 
Although we model data portraits as bipartite graphs between user interests (keywords) and user generated content (micro-posts), we do not focus on relations between words nor the text structure, and thus, we use \textit{word clouds} instead of any of the above mentioned techniques.

Visualization of micro-blog data covers a wide range of applications, including event monitoring \cite{dork2010visual,marcus2011twitinfo}, visual analysis~\cite{diakopoulos2010diamonds}, group content analysis \cite{archambault2011themecrowds}, and ego-networks \cite{2005-vizster}. 
We use \emph{data portraits}~\cite{donath2010data} to visually represent user profiles. Data portraits are \textit{``abstract representations of users' interaction history''} \cite{xiong1999peoplegarden}, 
and have been built for personal informatics systems~\cite{assogba2009mycrocosm}, Twitter profiles~\cite{lexigraphs} and discussion forums~\cite{xiong1999peoplegarden}, among others. 
Particularly, we borrow heavily from the work of \citet{viegas2006visualizing} in visualizing e-mail archives. E-mail archives are ``known datasets'', and users have expectations of what they will find on such portraits. If those expectations are fulfilled, the enjoyment of the application is greater. In addition, by reflecting on their past, users discover unknown patterns about themselves and their relationships. We rely on this  potential enjoyment and discovery to provide a positive experience for users, as a way to mitigate the cognitive dissonance that could be caused by our recommendations.

Graphical techniques used in recommendation contexts include controllable Venn-diagrams \cite{parra2015user}, network graphs \cite{gretarsson2010smallworlds,gretarsson2012topicnets}, \textit{dust and magnet} \cite{an2014sharing}, and compound graphs~\cite{gou2011sfviz}. These are  
 targeted at either expert users who know how to control such visualizations, or task-based systems.
Instead, similarly to the \textit{Hax} application \cite{savage2014visualizing}, we propose \emph{circle packing}, a hierarchical visualization technique~\cite{collins2003circle}, to create a casual and user friendly depiction of our generated recommendations, as it does not rely on user-controllability nor user expertise.
However, differently from \textit{Hax}, which builds a user interface to find audiences to broadcast information, our user interface allows users to find people to interact with.

\section{Methods and Design Rationale}
\label{sec:design}
For our work, we consider the micro-blogging platform Twitter, where users publish micro-posts known as \emph{tweets}, each having a maximum length of 140 characters. 
Each user can \emph{follow} other users, making their tweets available in his/her own \textit{timeline}.
Because of homophily \cite{mcpherson2001birds}, people tend to connect with like-minded individuals and do not think of the possible benefits of connecting with people of different (including opposing) opinions in sensitive issues. 
We design an intelligent system that recommends people to a target user, to study: 
(1) what factors affect the exploration and acceptance of recommendations built having politically diverse profiles into consideration, 
and (2) how to encourage unbiased behavior.

We introduce a data portrait design used in conjunction with a recommendation algorithm to recommend politically diverse profiles who share intermediary topics~\cite{graells2014people,DBLP:journals/corr/Graells-Garrido15a}. Although the data portrait is about each user's own data, we use it as a context to visually present the generated recommendations.

\begin{figure}[tb]
\centering
\includegraphics[width=\linewidth]{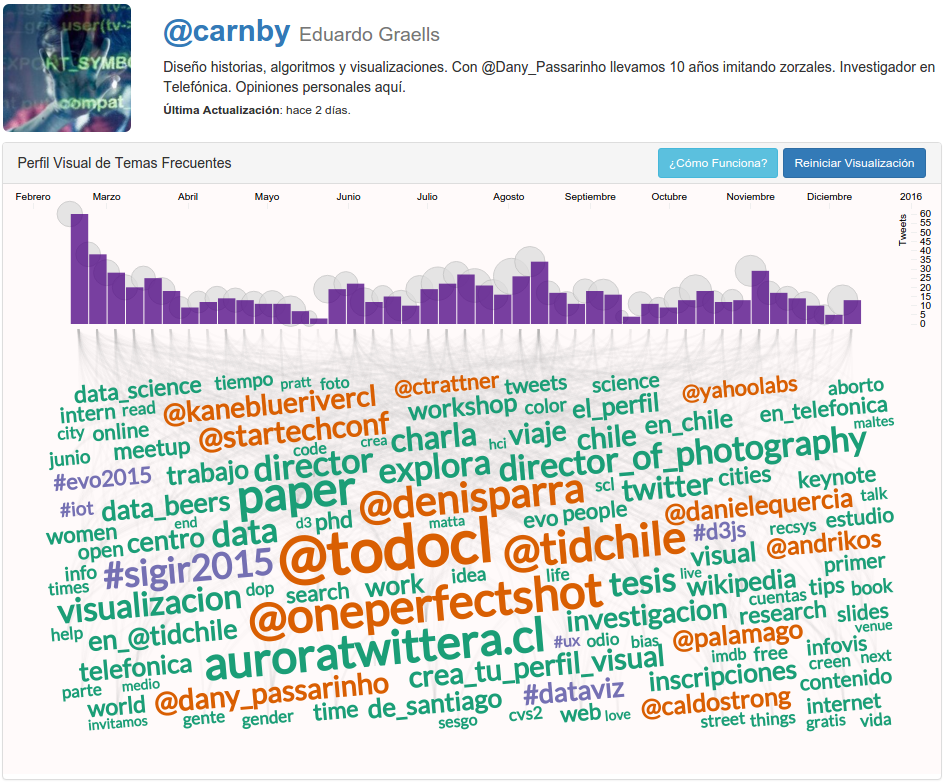}
\caption{Our data portrait design. In the image, the portrait of the Twitter account \textit{@carnby} (of Eduardo Graells-Garrido).}
\label{fig:data_portrait_new_design}
\end{figure}

\subsection{Portraying People's Data}
Data portraits \cite{donath2010data} can be used to create a self-image to present the target user. Based on the enjoyment and self-reflection experimented in ``known data'' scenarios~\cite{viegas2006visualizing}, our rationale is that we reinforce the non-conflicting interests of users when browsing their own profiles, while allowing to contextualize recommendations according to these interests. 

% \spara{Users Depiction}
Our proposed design is displayed in Figure~\ref{fig:data_portrait_new_design}. 
User interests are estimated by counting the frequencies of the n-grams (with $n \le 3$) of the words appearing in their timeline. 
The word cloud layout is based on \textit{Wordle}~\cite{viegas2009participatory}, allowing for a tighter yet flexible representation of words.\footnote{We use the implementation by Jason Davies available at \url{http://www.jasondavies.com/wordcloud/}.}
Note that common word clouds follow two patterns of rotation: random angle or 90\degree.
The first pattern makes reading the elements of the word cloud hard, whereas the second provides some sense of structure usually not present in the data. 
To promote a playful appearance, we decided to use rotated text. 
We used a fixed rotation for all words of -7\degree.  
This value was chosen arbitrarily after manual experimentation. As a way to compensate for the slight text rotation, to maintain readability we used a sans-serif font~\cite{rello2013good}.

The color coding of word cloud elements is based on the type of keyword. 
We consider three categories: \textit{hashtags} (\textit{\#7570b3}), \textit{mentions} (\textit{\#d95f02}) and \textit{regular words} (\textit{\#1b9e77}).
This palette is based on the color-blind friendly \textit{Set2} palette by \citet{harrower2003colorbrewer}, although we darkened the colors to provide a better contrast.
Each word has an invisible box that serves as clickable area, and as indicator when a particular word is highlighted when the box is visible.

\begin{figure*}[tb]
\centering
\includegraphics[width=0.33\linewidth]{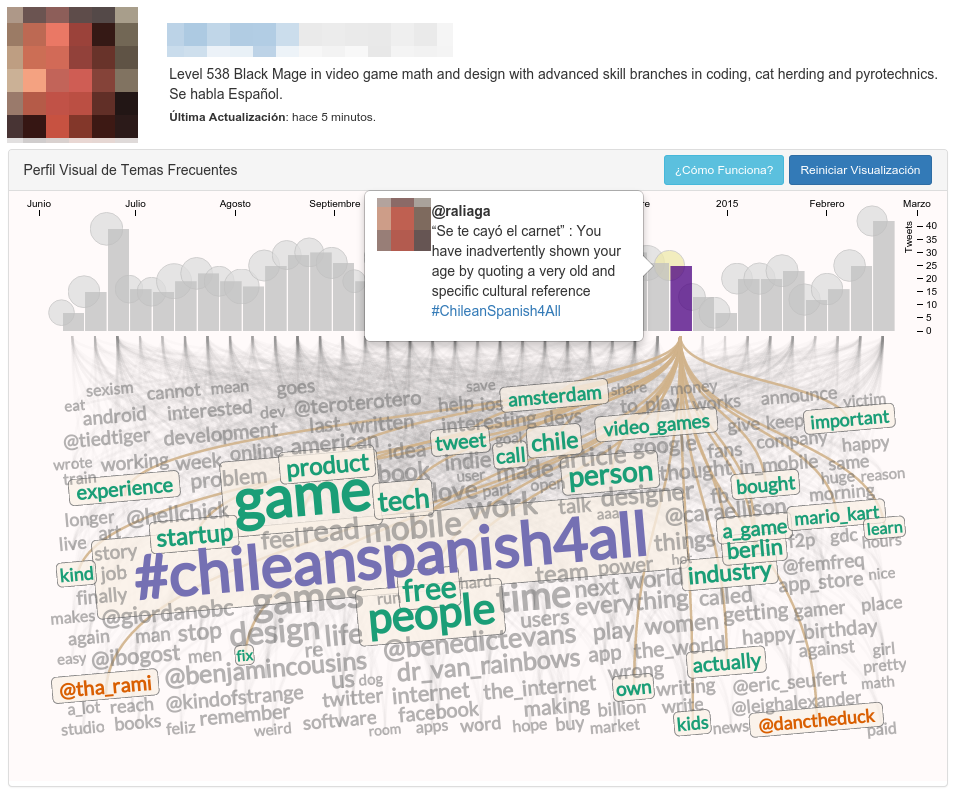}
\includegraphics[width=0.33\linewidth]{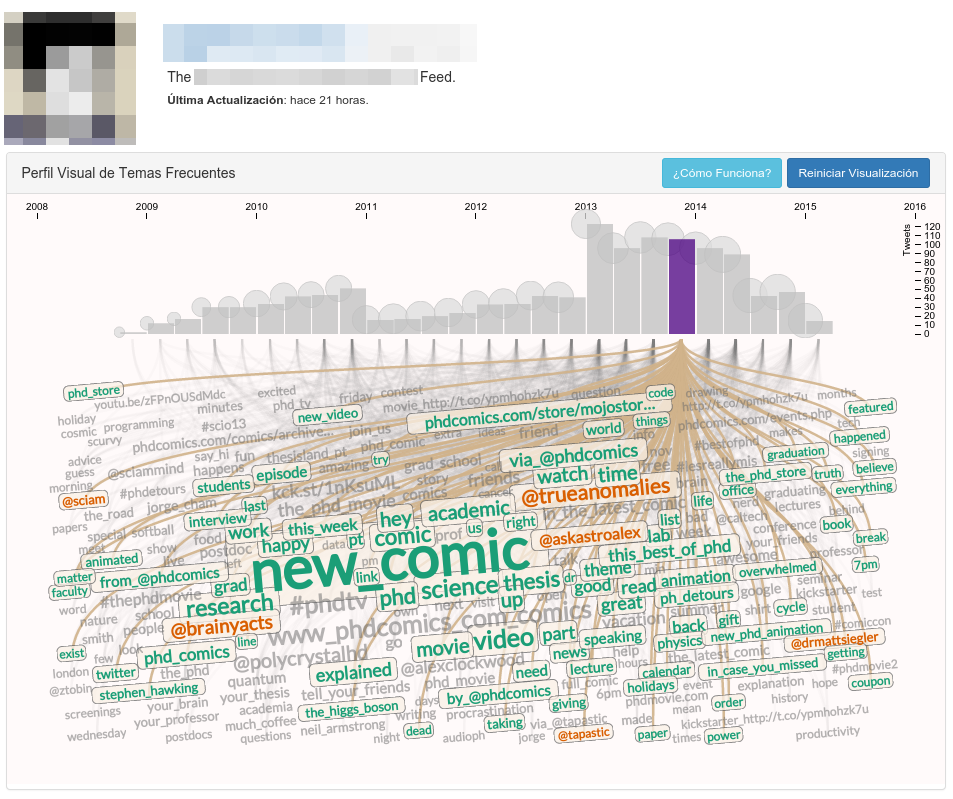}
\includegraphics[width=0.33\linewidth]{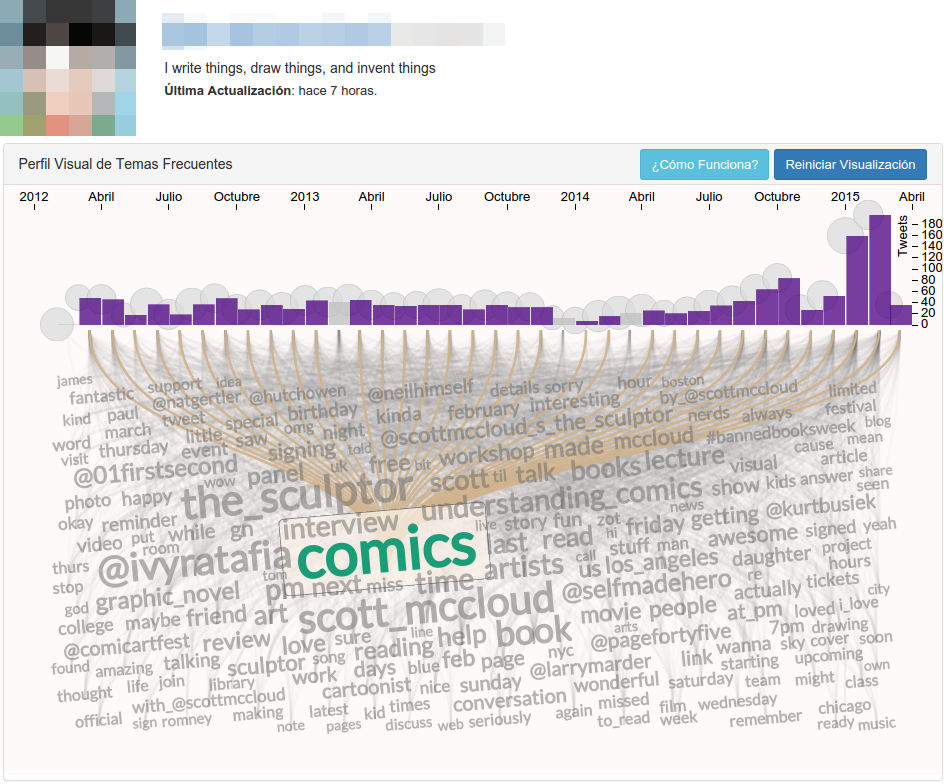}
\caption{State of the data portrait after several interactions. Left: a click on the histogram will display a tweet overlay, with links to all related keywords to the corresponding bin. Center: a click on the bin circle will deactivate the tweet overlay to ease exploration. Right: a click on a keyword will link related histogram bins. The source profile identities have been pixelated to maintain anonymity.}
\label{fig:data_portrait_new_design_interactions}
\end{figure*}

To provide temporal information,  we include a histogram of tweeting activity.
This histogram encodes the number of tweets published or retweeted during a given time window.
The number of bins is computed using Sturge's formula ($k = \lceil \log_2 n + 1 \rceil$) by the \textit{d3.js} library \cite{bostock2011d3}.
Each bin  is accompanied with a circle positioned on its upper-left corner, which is a turn-on/off switch for a tweet to be displayed in an overlay window. 
Although all circles are similar in size, their ratios vary slightly according to the popularity of the most popular tweet in the bin. Circles allow the user to select a bin regardless of its size, a feature particularly useful for time windows with low activity.

Finally, although users may want to change the background image of the data portrait, just like they can change the background on Twitter, we did not consider background customization to keep full control of aesthetics. Instead, we added their avatars and self-reported descriptions. 

\spara{Interactions and Component Linkage}
We link user interests and time (bins) using \textit{Bézier} curves.
The links are always visible, to make the structure behind the data portrait explicit to the user.
Those links are displayed in a non-highlighted state. To highlight links and change the state of the portrait, the following interactions are available:
\begin{itemize}
 \item When users click on a specific word, the corresponding bins are highlighted and connected through \textit{Bézier} curves (see Figure \ref{fig:data_portrait_new_design_interactions}, right).
 \item A click on a specific bin has two consequences:
 \begin{itemize}
    \item A tweet overlay is displayed with the most popular tweet in it (see Figure \ref{fig:data_portrait_new_design_interactions}, left). This tweet is context-dependent: if no word was selected before, it displays the overall most popular tweet; otherwise, it displays the most popular tweet relevant to the corresponding user interest. When a tweet is overlaid, the circle assigned to the current bin is highlighted.
    \item The words related to all tweets in the bin are highlighted (see Figure \ref{fig:data_portrait_new_design_interactions}, center).
 \end{itemize}
 \item When displaying a tweet overlay, if the user clicks its corresponding highlighted circle, the circle is desaturated (\ie, colored in gray), and the tweet overlay is hidden (see Figure \ref{fig:data_portrait_new_design_interactions}, center). 
\end{itemize}

The visualization can be reset by clicking on the \textit{``Reset Portrait''} button (\textit{``Reiniciar Visualización''}).
Additionally, we display a \textit{``How it Works?''} button (\textit{``¿Cómo Funciona?''}), which then displays a pop-up window with instructions.

\spara{Influence and Pilot Study}
Our proposed design is built upon the results of a pilot study \cite{graells2013data} and the design of \textit{Themail}~\cite{viegas2006visualizing}, a data portrait of e-mail conversations. In the pilot study, a data portrait was built following  an organic design to depict user profiles with recommendations injected inside it. The design was positively received by users, because displaying their interests allowed them to discover new things about themselves. 
It had, however, no impact on user behavior in terms of recommendations. In addition, several issues with the interface were raised, which were addressed in the new design (as shown in~Figure \ref{fig:data_portrait_new_design}), namely: 
consideration of time, a meaningful color palette for words, personalization of the data portrait, and better readability. 

\subsection{Recommending People Using Intermediary Topics}
Our proposed approach aims at recommending people that, while having a potentially distant position in political issues, have shared interests with the target user.
We do so by defining a proof-of-concept recommender algorithm. In this algorithm, the scoring of each recommendation is calculated by weighting two scores: 
1) a content-based distance using LDA;
and 2) an user similarity score based on \textit{intermediary topics}~\cite{graells2014people}, which we use as a proxy of shared interests. 
By weighting factors, this allows selecting a candidate that has intermediary topics with the target user, but is distant in terms of profile-wide topics as captured by LDA if the target user is politically vocal.
On the other hand, users that are similar in political content but with no intermediary topics would not be recommended to each other, as recommendations are based on shared interests characterized through intermediary topics.

These topics are obtained by running LDA over a corpus of user documents or \textit{microblogs} \cite{ramage2010characterizing}, and then creating a topic graph where nodes are topics. Two topics are connected through a weighted edge if both topics contribute content to at least one user.
Edge weights are based on the fraction of users that contribute to each edge.
Then, weighted closeness centrality is estimated in the graph, as a measure of how topics can make people closer.
Intermediary topics are defined as those in the top 50\% central topics, and have been found to be shared among a politically diverse set of people~\cite{DBLP:journals/corr/Graells-Garrido15a}.

\spara{Notations}
We represent a user $u$ as a vector:
\[\vec{u} = [p_{0}(u), p_{1}(u), \ldots, p_{k}(u)],\]
where $k$ is the number of latent topics, and $p_{i}(u)$ is $P(t_i \mid u)$ as defined by the LDA model for a topic $t_i$. 
Next, given two users, $u_1$ and $u_2$, we define their topical distance as the normalized \textit{Kullback-Leibler Symmetric Distance}~\cite{bigi2003using}:
\[KLD(u_1 \parallel u_2) = \sum_{i = 0}^{k} \{\vec{u}_{1}[i] - \vec{u}_{2}[i] \} \log \left( \frac{\vec{u}_{1}[i]}{\vec{u}_{2}[i]} \right).\]
To normalize a distance into the range $[0,1]$, given a set of distances, we divide each one by the maximum distance found. 

\spara{Similarity Considering Intermediary Topics}
We estimate \textit{similarity \wrt intermediary topics} and not with respect to all topics, as we want to recommend users who might not be close in distance terms (\eg, they could have ideological differences), but share intermediary topics. 
The set of intermediary topics for user $u$ is defined as:
\[IT(u) = \{ i: \vec{u}[i] \geq \varepsilon,~t_i \text{ is an intermediary topic}\},\]
where $\varepsilon$ is a threshold for topical significance, which depends on the context. 
For instance, the default value used in the \textit{gensim} library is 0.01 \cite{rehurek_lrec}.

We define similarity \wrt intermediary topics as the \textit{Jaccard Similarity} between two users:
\[JIT(u_1, u_2) = \frac{|IT(u_1) \cup IT(u_2)|}{|IT(u_1) \cap IT(u_2)|}.\]
Using this formula, when two users share all intermediary topics, $J(u_1, u_2) = 1$, and when users do not share any intermediary topic, $J(u_1, u_2) = 0$.

\spara{Algorithm Formalization}
Each candidate for recommendation is scored using a \textit{F-Score} \cite{baeza2011modern} of \textit{latent topical distance} and \textit{similarity \wrt intermediary topics}:
\[\text{score} = (1 + \gamma^2) \times \frac{\text{S}(u_1, u_2) \times (1 - \text{D}(u_1, u_2))}{\gamma^2 \times (1 - \text{D}(u_1, u_2)) + \text{S}(u_1, u_2)}.\]
where $S$ is similarity, $D$ is distance, and the balance factor $\gamma$ indicates the importance given to the distance in comparison to the importance given to similarity. For instance, $\gamma = 1$ gives equal importance to both factors, $\gamma = 0.5$ gives more importance to distance, and $\gamma = 2.0$ gives more importance to similarity.

Having estimated a measure of how close two users are, as well as how similar their sets of intermediary topics are, we can formalize our algorithm to recommend people with intermediary topics as follows. 
Given a target user $u$, a candidate set of recommendations $C$, a balancing factor $\gamma$, and the number of desired recommendations $n$,  we estimate the score for all candidates and return the top-$n$ scored candidates.

\subsection{Displaying Recommendations with Circle Packing}
The set of recommendations is displayed below the main data portrait as a separate unit, although both are clearly part of the same system.

% \spara{Recommendation Structure}
Our algorithm generates a list of recommended accounts to follow.
Each recommended account contains an avatar, a username, a biography, and a link to the full profile on Twitter. 
For each of those accounts we know the set of latents topics according to the LDA model.
To prepare recommendations for visualization, we cluster accounts based on their common latent topics.
We implemented a simple scheme, where two users are in the same cluster if their most contributing latent topic is the same. More complex clustering methods could be explored, but this is not the focus of this paper.

% \spara{Rationale}
Visual depictions have the potential to change how users perceive recommendations. 
First, visualization of social recommendations has been shown to increase user satisfaction~\cite{gretarsson2010smallworlds}. Second, explaining recommendations is important, as explanations increase user involvement and acceptance~\cite{herlocker2000explaining}.
Using visualization techniques to display recommendations allows us to depict the underlying structure behind them, hence providing an \textit{implicit} explanation. 
Conversely, when using text only, recommendations have to be explained in natural language, since something like \textit{``Topic 5''} is meaningless. Hence, visualization is a natural way to overcome this.

\begin{figure}[tb]
\centering
\includegraphics[width=\linewidth]{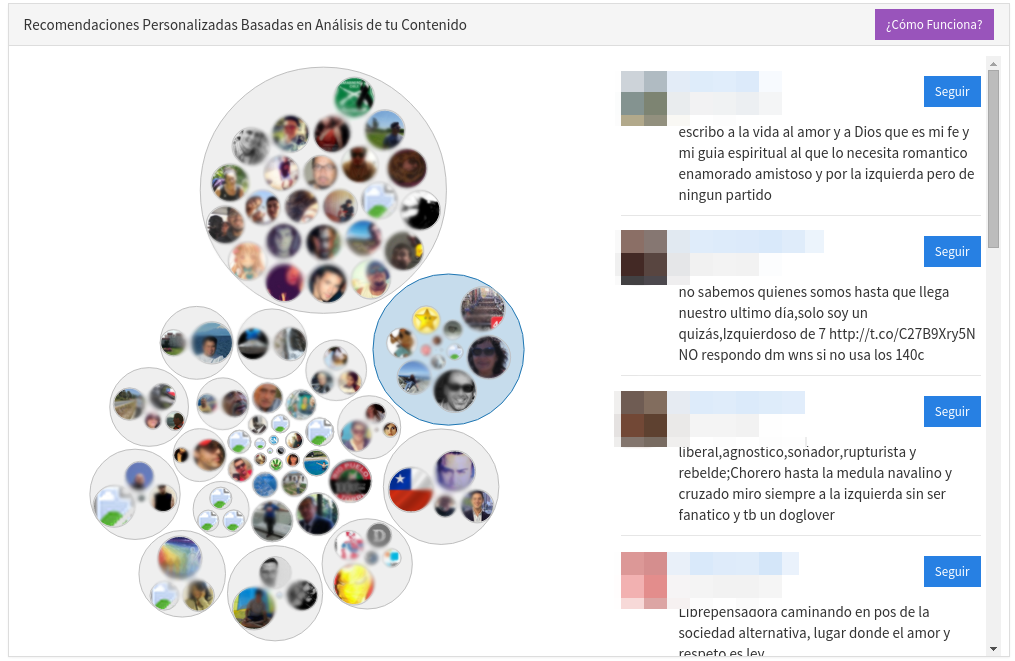}
\caption{Display of recommendations using Circle Packing. The recommended profile  identities have been pixelated to maintain anonymity.}
\label{fig:data_portrait_new_design_recommendations}
\end{figure}

% \spara{Hierarchical Visualization: Circle Packing}
We employ an \textit{enclosure diagram} built using \textit{Circle Packing}~\cite{collins2003circle} to display clustered recommendations.
Circle Packing (\textit{CP} hereafter) is a hierarchical visualization technique, where each node of the hierarchy is represented as a circle, with nesting according to the hierarchy. 
Even though this kind of diagram does not use effectively all the space available (for instance, unlike treemaps~\cite{bruls2000squarified}), CP \textit{``effectively reveals the hierarchy''}~\cite{heer2010tour}, and has an attractive organic appearance.
Additionally, circles maintain aspect ratio (unlike cells in a treemap), which is useful to display avatars of different sizes, as seen on previous applications in social query systems \cite{savage2014visualizing}.
An example visualization is shown in Figure~\ref{fig:data_portrait_new_design_recommendations}.

\spara{Interactivity}
Our design shows the CP visualization of the hierarchical structure of recommendations, but no actual detailed recommendation is shown at first. A message indicating that users can interact with the visualization is displayed instead.
Then, when users click on a cluster, the cluster is highlighted and a list on the right of the visualization displays a detailed list of the corresponding recommendations.
In the detailed list, each recommendation profile contains the account name, the full name of the recommended user, and the self-reported biography. The account name is linked to his/her profile on Twitter, and a ``Follow'' (\textit{``Seguir''}) button allows the user to directly follow the recommended account.

\section{Deployment ``In the Wild''}
\label{sec:in_the_wild}

We detail how we tested our platform, by deploying a proof-of-concept implementation in an uncontrolled setting~\cite{crabtree2013introduction}.

\subsection{Aurora Twittera: A Chilean News Aggregator Platform}
We implemented the visual designs using the \textit{d3.js} library~\cite{bostock2011d3}, and the recommendation algorithms using the LDA implementation from the \textit{gensim} library~\cite{rehurek_lrec}. 
We incorporated both, the data portrait design and the recommender system, in a Web platform named \textit{Aurora Twittera} \cite{auroratwittera} (available at \url{http://auroratwittera.cl}), \textit{AT} hereafter.
\textit{AT} is targeted at Chilean users (hence the \url{.cl} domain). It is a news aggregator that constantly crawls tweets about Chilean news, specially with respect to political events, but also news on other topics (\eg, sports, cultural events).

\subsection{Building Portraits}
In \textit{AT}, users could create their ``Visual Profiles'' (\textit{``Perfiles Visuales''}) by connecting their Twitter accounts with the site.
Since users had to log-in to the site to browse their portraits, we could gather rich interaction data, which we then analyzed.

After pressing the \textit{``Create Your Profile''} button, users were redirected to the Twitter website, which asked for login credentials and permission to modify their accounts.
We asked for these permissions to have a \textit{``Follow''} button next to each recommendation. %\footnote{Unfortunately, the text presented on the Twitter website made several users think that the site wanted to modify their public profiles, which likely negatively impacted the sign-up rate to our system.}
A scheduler service processed queued portraits, both, those newly created and those queued for update.
Tweets were downloaded using the Twitter API using the user credentials.
Then, we estimated the user interests according to the methodology described in the previous section and identified the intermediary topics using LDA. We considered the top-300 user interests according to frequency, a number chosen based on the width and height of the data portraits, which is suitable for a 1024x768 screen resolution.

From the up-to-date dataset of users in \textit{AT} we generated, every day, a list of candidate people who tweeted in the previous 48 hours, and for whom we estimate  their latent topics using the entire corpus of users who published tweets in those 48 hours.
This regularly updated list of candidates allowed us to present fresh recommendations to users.

\subsection{Promoting the System}
\textit{AT} has a social bot in Twitter, with username \textit{@todocl}.
The social bot \textit{@todocl} published tweets mentioning users when their portraits were ready (usually within less than one minute after sign-up), and every three days when their portraits were updated. Although portrait updates were daily, notification was limited to every three days to avoid spamming.
Additionally, to promote our system we performed a number of actions:
\begin{itemize}
 \item Created several \textit{demo portraits} for people to browse, and publicized them on \textit{@todocl}'s timeline.
The demo portraits were of popular user accounts. Sometimes the portrayed users, when being mentioned to notify them about the availability of their portraits, retweeted our announcements.
 \item Created a campaign on \url{http://ads.twitter.com} aimed at Chilean desktop users in Twitter who were active for at least one month. As result, 42,190 promoted tweets were displayed, with an engagement rate (as reported by Twitter) of 0.51\%.
 \item Added a ``Share my Profile'' button to the data portrait. When clicked, the system published a tweet from the portrayed user's account, inviting her/his followers to visit the data portrait.
\end{itemize}

The system was open to everyone. However, the user interface was available in Spanish only, and recommendations considered only Chilean users.

\section{Evaluation with Interaction Data}

Our hypothesis is that an indirect approach, through the mixture of data portraits and recommendations using intermediary topics, allows users to overcome the cognitive dissonance produced by exposure to potentially challenging information. To test this hypothesis our evaluation followed a \textit{between-groups} design. For each user who signed up on the system, a random pair of conditions $\left<UI, RecSys\right>$ was assigned. In both cases, user interface and recommender system, we considered a baseline condition in addition to our proposed one.

\begin{figure}[tb]
\centering
\includegraphics[width=\linewidth]{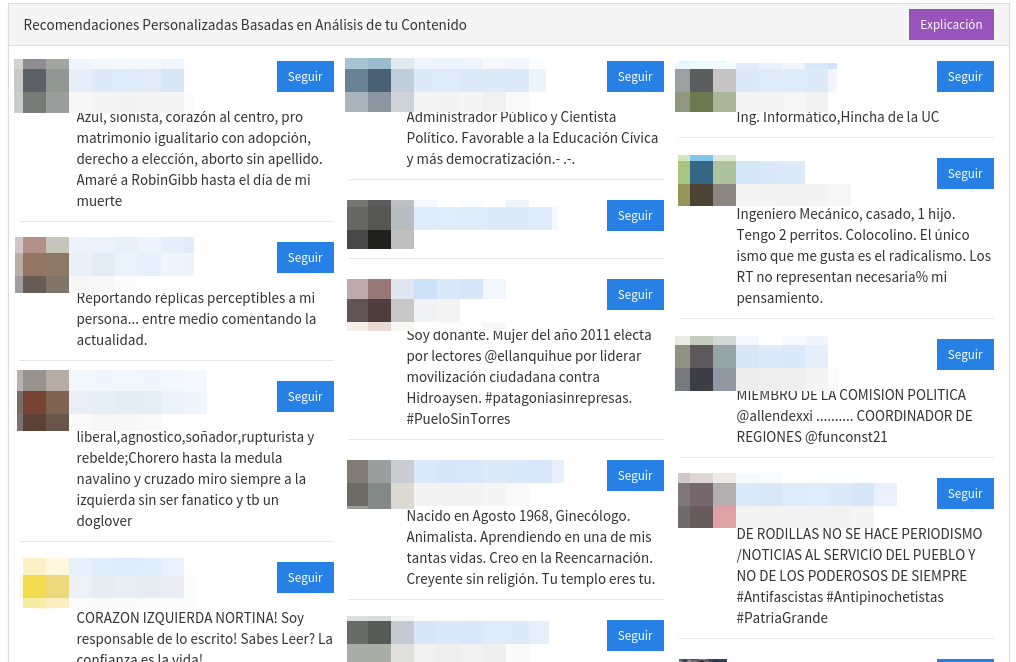} 
\caption{Baseline design of recommendations. The recommended profile  identities have been pixelated to maintain anonymity.}
\label{fig:data_portrait_baseline_recommendations}
\end{figure}

The \textit{User Interface} conditions were:
\begin{itemize}
 \item \textit{Baseline}: The baseline recommendation UI (see Figure \ref{fig:data_portrait_baseline_recommendations}), which displays recommendations in a similar way to current mainstream user interfaces.
 \item \textit{Circle Pack}: The visualization of recommendations using circle packing (see Figure \ref{fig:data_portrait_new_design_recommendations} Top).
\end{itemize}

The \textit{Recommender System} conditions were:
\begin{itemize}
 \item \textit{KLD}: recommendations generated using Kullback-Leibler Symmetric Distance only, \ie, considering the most similar users according to all LDA topics.
 \item \textit{IT}: recommendations generated using our proposed method based on intermediary topics (with balancing factor $\gamma = 1$).
\end{itemize}

In this way, we can compare whether using visualization influences how users explore recommendations, and whether including intermediary topics in the recommendation makes users behave in a less biased way.

%Each condition was randomly assigned to each user after receiving valid sign-in credentials from Twitter.

\begin{figure}[tb]
\centering
\includegraphics[width=\linewidth]{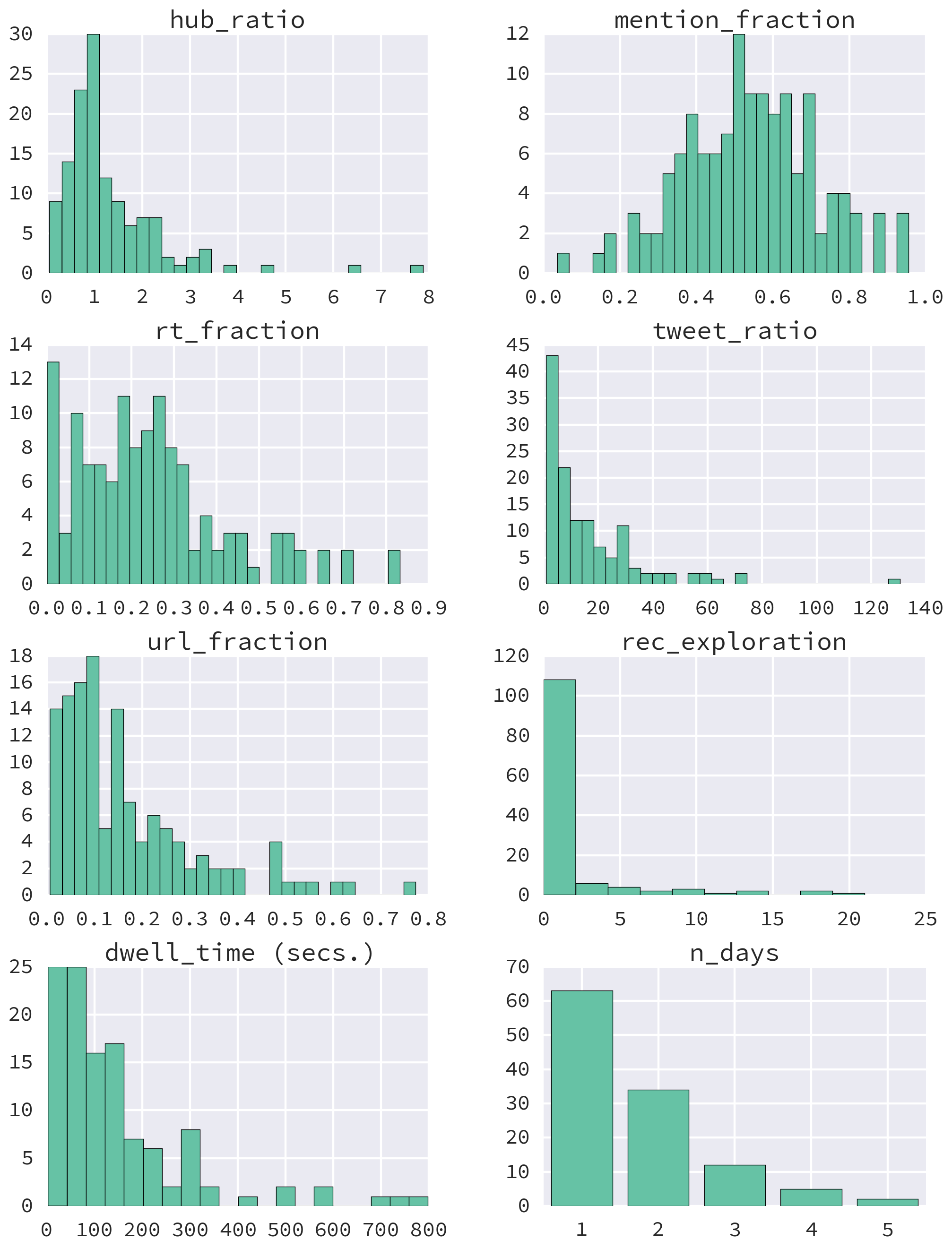}
\caption{The first five plots display the distribution of characteristics (independent variables) of portrayed users. The last three plots display the distributions of interaction data variables under study.}
\label{fig:data_portrait_interaction_data_users}
\end{figure}

% \spara{Independent Variables}
For each user, we had the following independent variables:
\begin{itemize}
 \item \textit{Political Content}: Its value is 1 if the list of the top-50 user interests has a non-empty intersection with a list of political keywords (including hashtags); the value is 0 if the intersection is empty. We considered this variable as binary, \eg, a user could use few political keywords, but if used very often, then it is arguably a user interested in politics.
 \item \textit{Hub Ratio}: Number of followed accounts divided by the number of followers, which measures the tendency of users to follow others based on their own popularity. 
 \item \textit{Mention Fraction}: Fraction of tweets that mention someone else (excluding retweets).
 \item \textit{RT Fraction}: Fraction of tweets that are retweets.
 \item \textit{Tweet Ratio}: Number of tweets per day, defined as total number of tweets published divided by the account age.
 \item \textit{URL Fraction}: Fraction of tweets that contain a URL (excluding retweets).
\end{itemize}

% \spara{Dependent Variables}
The interaction data we considered are explicit actions performed on the user interface:

\begin{itemize}
 \item \textit{Recommendation Exploration}: Number of clicks on elements in the user interface related to recommendations (\eg, click on a profile link or circle pack nodes).
 \item \textit{Recommendation Acceptance}: Whether the user accepted at least one recommendation.
\end{itemize}

Since \textit{AT} is a \textit{Casual InfoVis System} \cite{pousman2007casual}, users were not expected to perform specific tasks, nor instructed to do so. Although we compare interaction data between conditions to evaluate the differences in behavior, we still require a feedback mechanism to understand how users perceive the system. We analyze user engagement using \textit{implicit feedback}~\cite{lalmas2014measuring}, where  positive user engagement is used as our proxy of a positive perception of the system through the following variables:

\begin{itemize}
 \item \textit{Number of Days}: Number of days the user visited his/her data portrait.
 \item \textit{Dwell Time}: Time (in seconds) spent interacting with or browsing the data portrait with embedded recommendations.
\end{itemize}

% \spara{Statistical Model}
For our analysis, we consider the following factorial model:\footnote{Specified in R's formula syntax.}
\begin{equation*}
\begin{split}
Y & = C(\text{ui}) \times C(\text{recommendation}) \times C(\text{pol\_content})\\
  &+ \text{tweet\_ratio} + \text{hub\_ratio} + \text{RT\_fraction} \\
  &+ \text{URL\_fraction}+ \text{mention\_fraction},
\end{split}
\end{equation*}
where $C(X)$ creates dummy variables for the corresponding categories of the independent variable $X$, and $\times$ represents the independent factors and the interactions between them.\footnote{$A \times B = A + B + A * B.$}
We use this model for two types of regression: Negative Binomial (NB) generalized linear models, and logistic (logit) regression. NB is used for over-dispersed count data, and logit is used to model dichotomous outcomes in terms of probabilities.
In both models, if the statistical interactions between factors were found to be not significant, we analyzed the same model without interaction terms.

\subsection{Participants}
As our study focuses on Chilean users and Chilean politics, we discarded users whose self-reported Twitter location was not Chile, or whose IP address was not detected as Chilean by the GeoIP database.
We also discarded users whose interaction data was not reliable, \eg, having Javascript-blocking extensions in their browser. 
Lastly, we discarded users who spent less than 5 seconds on the site, and those whose tweet ratio was less than one.

As result, we have 129 valid portraits, created between 18 February and 17 March 2015.
For the recommendation conditions, 59 users were assigned to \textit{KLD}, and 70 to \textit{IT}.
With respect to the user interface, 59 users were assigned to the \textit{Baseline}, and 70 to the \textit{Circle Pack} condition. Finally, 69 users had political content in their portraits, and 60 did not.
The means of independent variables are: %\textit{account age}, 280 weeks; 
\textit{hub ratio}, 1.30; \textit{mention fraction}, 0.54; \textit{RT fraction}, 0.25; \textit{tweet ratio}, 16.54; and \textit{URL fraction}, 0.17. 
Figure \ref{fig:data_portrait_interaction_data_users} shows the distributions of these independent variables.

\begin{table*}[tb]
\centering
\footnotesize
\begin{tabulary}{\linewidth}{LLLLrrr}
\toprule
R\# & DV & IV & $\beta$ & Effect Size & 95\% C.I. & $p$-value \\
\midrule
R1 & Rec. Exploration & Intercept & $-2.104$ & --  &   $[-3.961,    -0.247]$      &   $0.026$ \\
R2 & Rec. Exploration & UI(CP) & $2.464$  & $11.750$ &   $[1.510,     3.417]$      &   $< 0.001$ \\
R3 & Rec. Exploration & REC(IT) & $-1.150$  & $-3.157$ &   $[-1.937,    -0.362]$      &   $0.004$ \\
R4 & Rec. Exploration & Tweet Ratio & $-0.030$ & $-1.031$ & $[-0.053,    -0.007]$      &   $0.012$ \\
R5 & Rec. Exploration & RT Fraction & $3.731$ & $41.721$ & $[1.468,     5.994]$      &   $0.001$ \\
% Accepted Recommendations
\midrule
R6 & P(Rec. Acceptance) & REC(IT) & $-4.383$ & $0.012$ &   $[-8.547,    -0.219]$      &   $0.039$ \\
R7 & P(Rec. Acceptance) & Pol. Content(True) &  $3.354$ & $28.617$ & $[0.061,     6.647]$      &   $0.046$ \\
R8 & P(Rec. Acceptance) & Mention Fraction &  $-12.079$ & $5.68 \times 10^6$  & $[-21.958,    -2.200]$      &   $0.017$ \\
R9 & P(Rec. Acceptance) & RT Fraction & $9.872$ & $19,380.060$ & $[1.026,    18.717]$      &   $0.029$ \\
% Dwell Time
\midrule
R10 & Dwell Time & Intercept & $4.825$  & -- &   $[3.971, 5.680]$      &   $< 0.001$ \\
R11 & Dwell Time & UI(CP), Pol. Content(True) and REC(IT) & $2.187$ & $8.908$ &   $[0.785, 3.590]$      &   $0.002$ \\
\bottomrule
\end{tabulary}
\caption{Regression Coefficients for the dependent variables under study. Only significant terms are shown for each regression.}
\label{table:new_portrait_variable_portrait_interactions}
\end{table*}

% \begin{figure}[tb]
% \centering
% \includegraphics[width=\linewidth]{img/interaction_data_results.png}
% \caption{Distributions of interaction data variables under study.}
% \label{fig:data_portrait_interaction_data_results}
% \end{figure}

\subsection{Regression Results}
The 129 portrayed users generated 1,707 interaction events. The following are the mean and max values found for each variable, as well as the results of the statistical analysis:

\begin{itemize}
\item \textit{Recommendation Exploration} events (mean $= 1.53$, max $= 21$): the first NB regression did not contain significant interactions. 
The NB model without interactions is (log-likelihood $= -169.59$; deviance $= 168.51$; $\chi^2 = 221$).

\item \textit{Recommendation Acceptance}: 5.42\% of participants accepted at least one recommendation. 
The first logit regression did not contain significant interactions. 
The logit model without interactions is (log-likelihood $= -14.67$, $p = 0.002$). 

\item \textit{Number of Days} (mean = 1.81, max = 8): none of the performed NB regressions reported significant terms.

\item \textit{Dwell Time}: 
We discarded the top decile from the analysis because some users left the browser window open (the maximum dwell time observed was 9 hours), 
leading to $N = 116$, with mean $= 147.66$, and max $= 798$ seconds. The NB model with interactions contained significant terms (log-likelihood $= -688.68$, deviance $= 95.89$, $\chi^2 = 80.5$).
\end{itemize}

Figure \ref{fig:data_portrait_interaction_data_users} shows the distributions of the independent variables. 
Table \ref{table:new_portrait_variable_portrait_interactions} displays the regression coefficients with a p-value smaller than 0.05, and the corresponding 95\% confidence intervals and p-value.
We refer to each result as R$i$. %A result can be a mixture of several coefficients in the presence of statistical interactions.
We discuss our results focusing first on recommendations, and then user engagement. 
As effect size for the logit regression coefficients, we consider the Odds-Ratio (OR) of each result. The OR is a measure of how associated a factor is to the outcome under analysis.
If the OR is greater than 1, then the presence of the factor is considered to be associated with the outcome. If the OR is lesser than 1, then the opposite association holds. An OR of 1 indicates no association.
The OR of each coefficient is defined as $\text{OR} = exp(\beta)$. 
In the case of the NB regression, the effect size of each coefficient is defined as $\text{ES} = exp(|\beta|) * \text{sign}(\beta)$. This ES means how much would the outcome increase (or decrease) with a one unit increase of the dependent variable.

\subsection{Recommendation Exploration}
We discuss the extent to which users explored the recommendations, and whether they accepted them  (users followed the recommended users). 
%For both, recommendation exploration and recommendation acceptance, we look at the effect sizes of the results of the regressions.

% \spara{Factors Associated with Exploration}
We found two positive effects that increase the tendency to explore recommendations. The strongest effect is the RT Fraction (R5, ES $= 41.721$). This can be interpreted as users who tend to retweet more, are more likely to explore recommendations because they are looking for sources to retweet.
The second positive effect is the usage of Circle Pack (R2, ES $= 11.750$). This ES indicates that exploration recommendations increase when users are exposed to CP, if all the other factors are held constant. This effect validates our design choice of using CP.

% \spara{Factors that Disencourage Exploration}
The negative effects are the usage of Intermediary Topics (IT) and Tweet Ratio. In the case of IT, users exposed to its generated recommendations decrease exploration (R3, ES $= -3.157$). This effect indicates that users tend to behave homophilically, probably because the recommendees' profile information can make their political leaning explicit (\ie, by using a politically-explicit avatar or self-reported description). However, its effect size is small in comparison to the positive effects found. 
Likewise, a one unit increase in Tweet Ratio decreases exploration (R4, ES $= -1.031$). This may indicate that users who tend to publish more tweets are less likely to explore recommendations because they are generating content--instead of looking for sources; they \textit{are} the sources.

\subsection{Recommendation Acceptance}
% \spara{Factors Associated with Recommendation Acceptance}
There are two significant positive effects. 
The strongest one is RT Fraction (R9, OR $= 19380.06$). 
We interpret this in concordance with previous interpretations, where users were looking for sources of information to retweet from.
The second positive effect is the presence of political content on a user portrait (R7, OR $= 28.617$). Hence, if the other variables are held constant, the odds of accepting a recommendation increase more than 28 times if the user is interested in politics. This aligns with results obtained in our first pilot study~\cite{graells2013data}.

% \spara{Factors that Disencourage Acceptance}
The negative effects are Intermediary Topics and Mention Fraction. In the case of IT (R6, OR $= 0.012$), this effect size confirms the homophilic behavior of the user population as already hinted by the previous result in recommendation exploration.
The other negative effect is Mention Fraction (R8, OR $= 5.68 \times 10^6$), which suggests that users with high fraction of mentions have almost negligible odds of accepting a recommendation. This suggests that users who already have a network of connections to interact with, do not need (or feel the need) to add new people in their networks.

\subsection{User Engagement}
We defined two variables related to user engagement: number of days that each user visited the site, and dwell time. Both variables allow to measure positive engagement with the site: someone who returns to the site in a different day may do so because s/he finds it useful, and someone who spends more time on the site, in a single session, may do so because s/he finds it interesting.

We observed that 45\% of participants returned to the site at least a second time on a different day. However, what causes this return cannot be explained by our regression model, implying that in terms of visits per day, all users were engaged equally.
With respect to dwell time, none of the standalone variables or main effects were found to be significant, nor the pairwise interactions. The only factor found significant is the triple interaction between Intermediary Topics, Circle Pack and Political Content (R11, ES $= 8.91$). 
In the NB model, this can be interpreted as follows: dwell time increases by 8.91 seconds when these three conditions are present, and all other factors are held constant.
Note that, if all possible interactions and main effects related to user interface, recommender system, and political content would be significant, there would still be an increase in dwell time by 2.06 seconds when those three conditions are present.
This could be because of increased satisfaction with the system \cite{fox2005evaluating}, or a deeper exploration of profiles. 

\subsection{Overview of Results}
Some results, positive and negative, support the motivation behind this work. On one hand, visualization of recommendations increased exploration. On the other hand, our proposed algorithm was not interesting for users, and was out-performed by a fully-homophilic baseline. Our results are not enough to fully support our hypothesis, although we found partial evidence in its favor: there is an effect of our mixed approach, but not for all users. 

It is well known that a majority of users are \textit{challenge-averse}~\cite{munson2010presenting}, so it could be expected that politically-engaged users exposed to our conditions would have had a negative (\eg, less engaging) experience. This was not the case -- users who were exposed to our proposed conditions and are politically-involved present a comparable (and even slightly more positive) experience when using the site. In the next section we discuss why this positive experience can be linked with conscious (unbiased) behavior.

\section{Discussion}
\label{sec:discussion}

\spara{Recommendation and Individual Differences}
To analyze individual differences, we focused on behavioral signals that could be extracted from user profiles, and applied a statistical model to find which ones influence user behavior with our system.
Some variables were not significant: \textit{hub ratio} (connectivity) and \textit{URL fraction} (type of content that is shared).
The significant variables were \textit{tweet ratio} (publishing behavior), \textit{mention fraction} (interaction with others), and \textit{RT Fraction} (information diffusion). 
Knowing that some users have the tendency to explore (or not) recommendations allows identifying the users who are more likely to benefit from them, and use simpler algorithms for those users who are not.

Furthermore, the user related variable that influenced recommendation acceptance was the presence of political content. 
In line with our motivation, arguably only politically-involved people are affected by selective exposure, in the sense that they look for political content, whereas non-politically involved people discard political content because of lack of interest instead of selective exposure. Not all users are interested in politics, therefore, not all users are interested in, nor need, political diversity on their timelines. 

Our visualization proposal was effective: when visualizing recommendations instead of using the baseline text interface, users' exploration of recommendation was equivalent, or even greater, than when recommendations were non-diverse. We used an aesthetically attractive design based on circle packing~\cite{collins2003circle}, which had the property of displaying part of the underlying structure in recommendations. 
In that aspect, we did not find a main effect of visualization in reducing homophilic behavior of recommendation acceptance.
In fact, homophilic behavior was confirmed: the non-diverse recommendation algorithm increased the likelihood of acceptance. However, due to the statistical interaction found, there is a potential usage of visualization when mixed with algorithms that recommend more diverse information. 

\spara{``Indirect'' Approaches and Unbiased Behavior}
As an indirect approach to exposing users to people of opposing views, we proposed a data portrait~\cite{donath2010data} context for users to explore their own content. These data portraits serve to display recommendations, where users could explicitly see how the system modeled their generated content, and implicitly understand the structure of the recommendations made to them.
Thus, when addressing the biased behavior of users, micro-blogging platforms may want to include other profile UIs, specially those based around data portraits~\cite{donath2010data} and \textit{Casual InfoVis} systems~\cite{pousman2007casual}.

Previous attempts at directly exposing users to opposing information had detrimental effects, given that a majority of users are \textit{challenge-averse} \cite{munson2010presenting}. 
For instance, in \textit{Opinion Space} \cite{faridani2010opinion}, although users did behave differently when using a visualization-based UI, their dwell times were not significantly different, and their political behavior was not less biased.
Our own results showed that politically-involved users under our proposed conditions did not behave in a less biased way, although they had slightly longer sessions. This is the key difference -- their extra-time browsing recommendations (from 2 to 8 seconds, with an average of 1.53 explored profiles) indicates that, whether they accepted or discarded recommendations, they took their time to explore/decide. 
We hypothesize that this is a \textit{conscious} decision-making process when exploring recommendations. Note that such consciousness cannot be achieved by direct approaches because they activate cognitive dissonance \cite{festinger1962theory}.

\spara{Limitations and Future Work}
In this work we used a proof-of-concept recommendation algorithm, based on the concept of intermediary topics. Although these topics have been shown to contain politically diverse profiles in the population under study~\cite{DBLP:journals/corr/Graells-Garrido15a}, there is a need to quantify in formal terms its differences with fully homophilic algorithms.
For instance, we focused on the political diversity of users provided by intermediary topics, but not on whether they actually had opposing views in sensitive issues or not.
Moreover, critics might rightly say that sentiment analysis could be used as proxy for opposed opinions instead of latent topics. However, these techniques are not language-agnostic and culture-independent, something relevant in our context (a population with Spanish as primary language), and thus this should be analyzed and tested as well.
Finally, our results indicate that our algorithm has lesser acceptance than baseline algorithms. Perhaps by considering more features expected by users, like network features~\cite{chen2009make}, results could be improved.

\section{Conclusions}
Behavioral and content differences influence how users perceive a \textit{Casual InfoVis} system~\cite{pousman2007casual}. 
In our research, while trying to understand how to encourage connecting with others who think differently, we found that being politically open, and the informational and interaction behavior of users, are important features that influence interaction.
This is important to be aware for the following reason. When designing systems for specific tasks, user characteristics can be assumed by visualization designers. However, in open systems like ours we cannot predict who will use the system nor their expertise level. In addition, in biased scenarios there does not seem to be a one-size-fits-all solution.
Because we presented an exploratory system, our evaluation was not a task-based one focused on algorithm/visualization efficiency.
Instead, we performed an ``in the wild'' evaluation~\cite{crabtree2013introduction}, where we focused on individual differences and user interaction with the interface, as well as user engagement metrics~\cite{lalmas2014measuring}. 
This allowed us to obtain deep insights on user behavior and exploratory styles.

To conclude, our results show that systems that aim at unbiasing user behavior should consider an indirect approach, which can be evaluated with engagement metrics in addition to context-specific metrics (\eg, recommendation acceptance). 
Furthermore, our work also shows that an unbiased behavior is not one that performs more actions (\eg, accepting more recommendations of people who thinks differently). 
On the contrary, it is one that allows users to avoid the cognitive heuristics that bias the activity, enabling them to make conscious choices.

\spara{Acknowledgments}
We are grateful to Daniele Quercia for inspiration and discussion.
We thank Shiri Dori-Hacohen and the anonymous reviewers for valuable feedback, and Andrés Lucero for help with the user studies.
This work was partially funded by Grant TIN2012-38741 (Understanding Social Media: An Integrated Data Mining Approach) of the Ministry of Economy and Competitiveness of Spain.

% \clearpage
\printbibliography

\end{document}